\documentclass{elsarticle}

\usepackage{hyperref}
\usepackage{fbox}
\usepackage[skins]{tcolorbox}

\usepackage{booktabs}  % publication-quality tables
\usepackage{eqparbox}  % sub-align table cells

\usepackage{soul,xcolor}
\usepackage{rotating}
\usepackage{threeparttable}
\usepackage{caption,booktabs,array}
\usepackage{dblfloatfix}
    {\endMakeFramed}

\newenvironment{frshaded*}{%
    \MakeFramed {\advance\hsize-\width \FrameRestore}}%
    {\endMakeFramed}
    
\newcommand{\rowgroup}[1]{\hspace{-1em}#1}

\newtcolorbox{mybox}{enhanced,sharp corners=all,colback=white,colframe=gray,toprule=0pt,bottomrule=0pt,leftrule=1pt,rightrule=1pt,overlay={
    \draw[gray,line width=1pt] (frame.north west) -- ++(2cm,0pt);
    \draw[gray,line width=1pt] (frame.south east) -- ++(-2cm,0pt);
}}

\begin{document}

\setstcolor{red}

\begin{frontmatter}

\title{Industry-Academia Research Collaboration in Software Engineering: The Certus Model}

%% Group authors per affiliation:
\author{Dusica Marijan\corref{cor1}}
\address{Simula Research Laboratory, Norway}
\ead{dusica@simula.no}

\author{Arnaud Gotlieb}
\address{Simula Research Laboratory, Norway}

\cortext[cor1]{Corresponding author}

\begin{abstract}
\textbf{Context}: Research collaborations between software engineering industry and academia can provide significant benefits to both sides, including improved innovation capacity for industry, and real-world environment for motivating and validating research ideas. However, building scalable and effective research collaborations in software engineering is known to be challenging. While such challenges can be varied and many, in this paper we focus on the challenges of achieving participative knowledge creation supported by active dialog between industry and academia and continuous commitment to joint problem solving. 
\textbf{Objective}: This paper aims to understand what are the elements of a successful industry-academia collaboration that enable the culture of participative knowledge creation. 
\textbf{Method}: We conducted participant observation collecting qualitative data spanning 8 years of collaborative research between a software engineering research group on software V\&V and the Norwegian IT sector. The collected data was analyzed and synthesized into a practical collaboration model, named the Certus Model. \textbf{Results}: The model is structured in seven phases, describing activities from setting up research projects to the exploitation of research results. As such, the Certus model advances other collaborations models from literature by delineating different phases covering the complete life cycle of participative research knowledge creation. \textbf{Conclusion}: The Certus model describes the elements of a research collaboration process between researchers and practitioners in software engineering, grounded on the principles of research knowledge co-creation and continuous commitment to joint problem solving. The model can be applied and tested in other contexts where it may be adapted to the local context through experimentation.
\end{abstract}

\begin{keyword}
Software engineering\sep Industry-academia collaboration\sep Research collaboration\sep Research knowledge co-creation\sep Collaboration model\sep Technology transfer\sep Knowledge transfer\sep Research exploitation\sep Research-based innovation 
\end{keyword}

\end{frontmatter}

%\linenumbers

\section{Introduction}
Industry-academia collaborations in software engineering (SE) have been an important discussion topic in the SE community for many years. This is because research collaborations between SE industry and academia, if managed successfully, can provide significant benefits to both sides, such as improved innovation capacity for industry \cite{Wirsich2016}, and real-world environment for validating research outputs \cite{Perkmann2010} as well as building new areas of research expertise for academia \cite{Lamprecht2012}. However, as observed by Bosch \cite{Bosch2014}, building scalable and effective research collaborations between industry and academia in SE is known to be challenging. Chimalakonda \cite{Chimalakonda2015} summarizes some of the challenges that both sides of such collaborative projects face, observing that most practitioners' have perception that researchers work on dated or futuristic theoretical challenges which are divorced from industrial practice, while researchers believe that practitioners are looking for quick fixes instead of using systematic methods. Similarly, it is observed that researchers are more interested in proposing new techniques and tools, focusing on technical novelty, while practitioners would appreciate solutions that work in their context, regardless of novelty. Runeson \cite{Runeson2014a} specifically analyzes the time horizon aspects of long term industry-academia projects, which are generally shorter in industry compared to the academic perspective, creating risks for friction and frustration on both sides of the collaboration. Discussing key challenges of industry-academia collaboration, Barroca \cite{Barroca2015} points out that timeliness is very important for industry and less important for academia. She further stresses that research outputs need to be crafted in different ways to be relevant to different stakeholders, that researchers focus on rigour while industry focuses on pragmatic approaches, and that academics tend to report research findings in formats not easily accessible to practitioners. Ivanov \cite{Ivanov2017} provides more examples of gaps between research and practice, collected from a survey, concluding that there is a wide gap between the problems that practitioners face and the research topics discussed at academic conferences.

In an attempt to reduce such a state of practice, there have been numerous efforts to understand the collaboration challenges and ways to make collaboration projects more successful. For example, the IMPACT project of ACM SIGSOFT tried to measure the impact of SE research on practice over 6 years \cite{Osterweil2008}\cite{Emmerich2005}, pointing out that the transfer of ideas from academia to industry is challenging and can take roughly 15 to 20 years. 
At ICSE 2011, the participants of the panel "What industry wants from research" \cite{Icse2011} acknowledged the industry-academia collaboration gap in SE and tried to identify key challenges of bringing theory and practice closer together. Some of the panelists suggested that to reduce the gap between industry and academia in SE, industry should communicate their real problems and should share their case studies, while researchers should improve usability of the tools and methods developed, and consider how to improve value, especially earning. In his ICSE 2012 keynote, Briand \cite{Briand2012} called attention to the need for inductive research, which starts from specific observations in real settings, as well as more thorough understanding of practice, along with closer interactions with practitioners in the definition of research problems. This idea has been further emphasized in \cite{Briand2017} by advocating the case for context-driven research. According to Briand, instead of devising univeral solutions that generalize to any context, context-driven research strives for practicality and scalability satisfying the constraints in the given context. In spite of all these efforts, 20 years later, there still exist significant challenges for SE researchers and practitioners standing in the way of establishing and successfully running collaborative projects \cite{Bern2018} \cite{Ivanov2017} \cite{Tang2017} \cite{Garousi2018}. At the same time, we observe the need to establish more successful collaborations in SE \cite{Basili2018} \cite{Carver2018}.   

In this paper, we focus on industry-academia collaboration challenges pertaining to research knowledge co-creation, where industry and academia collaboratively work on problem definition and solving, with continuous dialog and goal alignment to create artifacts of value for both sides.

\textbf{\textit{Research Approach}}. The paper addresses the research question \textit{What elements are needed for an industry-academia collaboration to support research knowledge co-creation in SE?} To answer this question, we used the qualitative research method of participant observation for data collection. The data was collected continuously in a longitudinal study spanning 8 years of the Certus Center existence. The data consists of field notes taken during day-to-day discussions among the project participants, as well workshops and focus group meetings dedicated to discussing the Certus collaboration setup. The data also includes several hundred emails exchanged between the project participants discussing the collaboration. The data was analyzed to derive findings on the collaboration which were validated by the project participants using the member checking method. The findings were used to build the Certus model which was iteratively refined throughout the collaboration as more observation data and participants' feedback became available.

\textbf{\textit{Contributions}}. 
First, this paper contributes to the body of evidence on bridging the industry-academia collaboration gap. Second, it presents a practical "good practice" model aimed at improving research collaborations between industry and academia based on research knowledge co-creation. We argue that the co-creation thinking has the potential to prevent the drop of interest and commitment between collaborating partners, by enforcing continuous commitment and goal alignment, resulting in more effective value creation based on collective creativity. Third, the model introduces seven elements of industry-academia collaboration in SE. At its core, each element assumes continuous interaction, commitment and alignment between industry and academia, thus supporting research knowledge co-creation.

\section{Background}
In existing work on industry-academia collaboration, there are: (i) reviews and surveys on challenges and best practices for industry-academia collaboration, (ii) case studies reporting success factors and lessons learned on the subject of industry-academia collaboration, and (iii) existing collaboration models, some of which inspired our work.

\subsection{Challenges and Best Practices for Collaboration} 
Garousi \cite{Garousi2016} \cite{Garousi2017a} reviews a number of projects describing industry-academia collaboration in SE and synthesizes a list of patterns, anti-patterns, and challenges in collaboration. The challenges are categorized according to the themes including the lack of research relevance, lack of training, experience and skills, lack of interest and commitment, mismatch between industry and academia in terms of terminology and communication, human and organizational factors contributig to resistance to change, management related issues, contractual and privacy concerns in connection with trust and respect. The study further identifies the set of best practices addressing each of these challenges. Pertuze \cite{Pertuze2010} surveys more than 100 projects at multinational companies that engage in research collaborations with academia, reporting seven guidelines for companies to increase collaboration success. The guidelines include defining the project’s strategic context as part of the selection process, selecting boundary-spanning project managers, sharing with the university research team the vision of how the collaboration can help the company, investing in long-term relationships, establishing strong communication linkage with the university team, building broad awareness of the project within the company, and supporting the work internally both during the contract and after, until the research can be exploited. 

\subsection{Collaboration Case Studies}
Wohlin \cite{Wohlin2011} analyzes success factors for two industry-academia collaboration projects in SE in Sweden and Australia, concluding that the industrial side of collaboration is the key element for successful collaboration. He also discusses top ten challenges for industry-academia collaborations, some of which include trust and respect, roles and goals, and knowledge exchange instead of technology transfer. In addition, he proposes five levels of closeness between industry and academia \cite{Wohlin2013}, ranging from less to more close: not in touch, hearsay, sales pitch, offline, and one team. Petersen \cite{Petersen2014} describes experience from two case studies of industry-academia collaboration based on action research principles. In these studies, close collaboration and co-production with industry is the key to make refinements and build solid understanding of the context. Sandberg \cite{Sandberg2011} summarizes lessons learned in a long term industry-academia collaboration project, describing success factors enabling research activities and results, including management engagement, network access, collaborator match, continuity, communication ability, need orientation, industry goal alignment, deployment impact, industry benefits, and innovativeness. Sjoo \cite{Sjoo2019} applies a systematic literature review to identify enabling factors for collaborative innovation between industry and academia according to timeframe for collaboration. In the short term, such enabling factors include resources, organizational capacities and Intellectual Propetry Rights (IPR), boundary-spanning functions such as related ongoing activities. In the medium term, the factors include collaborative experience, and in the long-term, culture, status of actors and environmental factors. Dittrich \cite{Dittrich2008} reports on the experience of applying Cooperative Method Development (CMD) approach to research projects conducted in collaboration with industry. CMD consists of five guidelines, the first of which, action research, makes an overlap with our work. Action research, as described in the paper, consists of three phases: understanding, deliberating change, and implementation and evaluation of improvement. Barroca \cite{Barroca2015} presents lessons learned from applying a dedicated research collaboration model consisting of three phases: collaboration kick-off, investigation, implementation and evaluation, to three agile project with industry. Lessons learned from these projects are that building trust is important and can be achieved by giving feedback early and frequently, written agreements are useful for a running project smoothly, flexibility is important to accommodate changes, there should be outputs tailored to different audiences, cash investment from industry helps show commitment, relevant research expertise is necessary, regular contacts with industry help prevent drop out, and too much choice and information can be overwhelming. Mathiassen \cite{Mathiassen2012} reports experiences from a three-year industry-academia collaboration reflecting on research goals, approaches, and results. He emphasizes the importance of a combined approach based on action research, experiments and conventional practice to strike a useful balance between relevance and rigor in practice research.

\subsection{Existing Collaboration Models}
There exists literature on theoretical aspects of industry-academia collaboration \cite{Lin2017} \cite{Huang2017} \cite{Banal2013}. However, this is theoretical work originating from economics and research policy fields, and not the SE field. Several authors propose models for structuring industry-academia collaborations in SE. Sjoo \cite{Sjoo2019} presents a logical model that considers a number of hypothesis about the collaboration, and as such, the model also belongs to the body of theoretical work. In the SE field, Runeson \cite{Runeson2014} presents the 4+1 model of industry-academia collaboration taking five views on the collaboration process: time, space, activity, domain, and scenario binding the other four views. However, this is an architectural model, describing relationships between different components in the collaboration, not considering process aspects. Garousi proposes a process model for industry-academia collaboration, in a form of a context diagram \cite{Garousi2019}. This model focuses on the process aspect of collaboration and on cause-effect relationship in collaboration, consisting of inception, planning, operational, and transition phases. However, this is a theoretical model not describing the use of the model in context. Unlike these models, the model presented in this paper is a practical collaboration model, derived from our eight-year experience of running a collaborative research project with the SE industry. The work that is closest to ours is the model for technology transfer introduced by Gorschek \cite{Gorschek2006}. This model, which inspired our work, is inspired by the Pfleeger's model of technology transfer \cite{Pfleeger1999}. The Gorschek's model proposes a seven-step collaboration process, with a high-level view on each step. The seven steps include: basing research agenda on industry needs, problem formulation, candidate solution formulation, lab validation, static validation, dynamic validation, and solution release. We considered this model when thinking about how to facilitate industry-academia collaboration in our context. While we found the model very useful we also realized that we need a more detailed collaboration framework, which encompasses the full set of different dimensions of the collaboration process. 
This was especially true with respect to the roles and responsibilities of different participants, team composition, building and keeping a dialog between researchers and practitioners, promoting commitment and participative knowledge generation, expectation management, as well as considerations for commercialization of research results, which are some of the differing aspects of the model presented in this paper compared to the Gorschek's model. Thus, the gap in knowledge this research intends to fill is to empirically describe detailed stages of participative knowledge generation between industry and academia, including roles of participants, principles of interaction and means of building and sustaining continuous commitment and alignment.

\subsubsection{Comparison with Other Research\&Innovation Centers}
Furthermore, we have analyzed two Research\&Innovation centers whose industry-academia collaboration models influenced our work, to pinpoint some similarities and differences in their models compared to ours. They are SnT Center in Luxembourg and CRIM Center in Montreal. SnT \footnote{https://wwwen.uni.lu/snt} is an Interdisciplinary Centre for Security, Reliability and Trust. SnT conducts internationally competitive research in ICT, aimed at creating socio-economic impact. In addition to long-term, high-risk research, SnT engages in demand driven collaborative projects with industry and the public sector, through its Partnership Program. In that sense, SnT is very similar to Certus. Furthermore, V\&V Lab for Software Verification and Validation (V\&V) at SnT, which was lead by Prof Lionel Briand, has a very similar thematic area to Certus. Therefore, we found it relevant to draw a line between the V\&V lab at SnT and Certus. The data for the comparative analysis between Certus and the V\&V lab at SnT is derived from numerous discussions held with Lionel Briand, who was briefly a Certus member at the beginning. There are two particular concepts that Certus adopted and that can be atributed to Lionel Briand. The first is the concept of a context-driven research \cite{Briand2017}, which means that in order to create impact for industry, research needs to focus on problems defined in collaboration with industrial partners. Through experimentation, Certus has realized the importance of it and has adopted context-driven research as one of pillars of its collaboration model. Another concept and further similarity between the two labs is the notion of an industry champion. Having active industry champions in some of the Certus projects proved an important element of a successful industry-academia collaboration.

CRIM \footnote{https://www.crim.ca/en} is an applied research and expertise center in ICT. It aims to contribute to scientific advancement and to help organizations be more competitive through the development of innovative technology and the transfer of know-how. CRIM has one of the largest networks of IT companies in Quebec, which makes a rich innovation ecosystem with different actors and resources needed for the center to achieve its mission. After visiting CRIM and discussing with its members, it became clear that such an ecosystem with high innovation and absorptive capacity is a crucial factor facilitating a collaboration with industry. Certus on the other hand, operated in a less-rich innovation ecosystem compared to that of CRIM, which made performing research-based innovation more challenging. The companies we collaborated with did not have such high levels of absorptive capacity, which is the ability to appropriate research- and innovation-based results. Motivated by CRIM, we realized that in order to foster the scientific and innovation excellence of the center, we must build a dedicated collaboration model which will facilitate knowledge creation with and exploitation by industry, resulting in improved absorptive capacity for our industry partners, which in turn will create richer innovation ecosystem for Certus to operate in.

There are many other successful Research\&Innovation centers in SE in the world, for example the Fraunhofer Center for Experimental
Software Engineering in Maryland, known for applied research and the CREST Center for Research on Evolution, Search and Testing at the University College London UK, to name a few. However, there is no publicly available information on the details of their collaboration models with industry, to make it possible for us to compare.

\section{Case Study Context}
This section describes the context of the industry-academia collaboration, which was the basis for developing the Certus collaboration model. 

Certus is a Norwegian eight-year research-based innovation center that started in 2011 and concluded in 2019. It engaged a SE research group at Simula Research Laboratory (SRL) \footnote{www.simula.no,  Simula was founded in 2001.}, acting as a host institution, five large for-profit companies: Cisco Systems Norway (CIS), ABB Robotics (ABB), FMC Technologies (FMC), Kongsberg Maritime (KM), Esito (ESI) and two public sector institutions: the Norwegian Customs and Toll (TOL) and the Cancer Registry of Norway (CRN). For the sake of simplicity, we refer to both the collaboration between Simula and industrial companies and between Simula and public sector institutions as industry-academia collaboration. Similarly, we jointly refer to all Certus partners as industry partners. Simula is a publicly-funded Norwegian research institution whose mission is to create knowledge about fundamental scientific challenges that are of genuine value for the industry and society. An overview of the Certus partners and Simula, along with their domains, motivation for the collaboration, dates of participation in Certus and main outputs produced is presented in Table 1.

\begin{sidewaystable}
\centering
\scriptsize
\caption{Certus Partners: domains, motivations for collaboration and outputs created.}
\begin{threeparttable}
\begin{tabular}{p{0.9cm} | p{0.8cm} | p{1.3cm} | p{2.5cm} | p{3.9cm} | p{7.7cm}}
\hline
\textbf{Partner} & \textbf{Type} & \textbf{Years} & \textbf{Domain} & \textbf{Motivation} & \textbf{Main outputs}\\
\hline \hline
SRL & RI\tnote{1} & 2011-2019 & Research and innovation on software V\&V & Develop, validate, and deploy novel techniques for software V\&V. Publish high quality research papers. Educate PhD students & 5 prototype tools developed and deployed in industry contexts. Around 100 journal/conference publications. 1 patent application. 1 open source tool released. 5 PhDs completed. 1 industrial PhD supervision.\\
\hline
CIS & LI\tnote{2} & 2011-2019 & Video communication software & Reduce regression testing costs & Prototype tool for test optimization that reduces test cost. A number of publications co-authored with researchers\\
\hline
ABB & LI & 2014-2019 & Robot control software & Reduce the cost of test selection and scheduling & Prototype tool for test prioritization and scheduling. 1 industrial PhD completion. A number of publications co-authored with researchers\\
\hline
FMC & LI & 2011-2013 & Subsea control systems& Reduce the costs of system (re)configuration. & Prototype tool for configuration of the integrated control system. A number of publications co-authored with researchers\\
\hline
KM & LI & 2011-2018 & Subsea control systems & Improve the cost-effectiveness of safety analysis and certification & Prototype tool for evidence evolution management. Prototype tool for history-based recommendations for testing\\
\hline
ESI & SME\tnote{3} & 2011-2019 & Software tools for domain driven development & Create new market opportunities by exploiting  research-based innovation results. & Business opportunity to commercialize the tool for regression testing of data-intensive systems developed in collaboration with SRL and TOL partners\\
\hline
TOL & PS\tnote{4} & 2011-2017 & Customs accounting & Automate non-regression testing. Improve the quality of tests & Prototype tool for regression testing of data-intensive systems. Prototype tool for visualizing test data interaction coverage. A number of publications co-authored with researchers\\
\hline
CRN & PS & 2015-2019 & Cancer registry system & Transform a manual system into an ICT-based system, while ensuring system quality & Methodology for testing the integration of the CRN technical platform for screening with other services\\
\hline
\end{tabular}
\begin{tablenotes}\footnotesize
\item[1] Research Institution, $^2$ Large Industry, $^3$ Small to Medium Enterprise, $^4$ Public Sector
\end{tablenotes}
\end{threeparttable}
\end{sidewaystable} 

Certus has conducted research in the field of software testing, verification and validation, targeting the following types of systems: (i) real-time embedded software systems, (ii) highly-configurable software systems, and (iii) data-intensive software systems. The mandate of the Certus Center was to conduct industry-relevant research, where researchers and practitioners work collaboratively to define problems of mutual interest to address and create value for both sides. The Certus collaboration excluded consultancy activities, which are characterized by lesser degree of freedom to define problems to work on.

The Certus Center was funded by the Norwegian Research Council, the host institution, and the industry partners (in-kind), for eight years, with around 21 million NOK per year. The research institution partner has committed six senior researchers, six PhD students, and one research engineer to the project. All original Certus projects ended in 2019, due to the ending of funding. However, some of the partners continued the collaboration through new projects. ABB and SRL continued collaboration in a new project that is adjacent to the project of testing robot control software, addressed in Certus. CRN and SRL continued to work together in a new project that is related to testing of the cancer registry system, which was addressed in Certus.

\subsection{Project Setup}
The Certus Center was structured into 6 scientific projects, each dealing with a specific thematic area of a broader field of software V\&V, and 4 governing projects, dealing with overall project management, research exploitation and dissemination. An overview of the Certus projects is presented in Table 2.

Scientific projects addressing V\&V challenges common to several partners crosscut multiple partner domains where there is interest. In this way, we achieved the increase of interaction and communication between the partners, relating to concrete challenges and their technical solutions, which allowed for common areas of interest to arise and create synergies.

Among the governing projects, the \textit{Management} project involved the activities such as projects' financial governance, reporting to the funding body, definition of the Certus annual work plan (describing planned activities for all projects), making strategic decisions of involving new partners, allocating resources to projects, liaising between the partners in case of conflicts, monitoring the projects' progress, making decisions to initiate market research activities for promising technologies produced in Certus, and running administrative governance (collecting deliverables, minutes from meetings, reports from mid-way evaluations). The \textit{Research Exploitation} project dealt with collecting and analysing the data for the Certus collaboration model development, the development and updating of the model. The \textit{Training} project was providing courses to train industry professional with the state of the art in software V\&V. The \textit{Dissemination and Communication} project was making ongoing efforts throughout the term of the Certus Center to disseminate research results and other relevant information regarding the Certus operation in popularized, scientific and industrialized form.

\begin{table}[h]
\scriptsize
\caption{Overview of scientific and governing projects, and involvement of the partners across the projects.}
\centering
\begin{tabular}{>{\quad}lc}
\toprule
& \multicolumn{1}{c}{Participants}  \\
\midrule
\rowgroup{\textit{Scientific projects}} \\
Model-Based Engineering for Highly Configurable Systems & SRL, FMC, KM  \\
Safety Analysis and Certification of Embedded Systems & SRL, FMC, KM  \\
Testing of Real-Time Embedded Systems & SRL, CIS, ABB  \\
Testing of Data-intensive Systems & SRL, TOL, ESI, CRN  \\
Smarter V\&V of Evolving Software Systems & SRL, CIS, FMC  \\
Data-Driven Predictive Maintenance for Software Systems & SRL, CIS, ABB \\
\midrule
\rowgroup{\textit{Governing projects}} \\
Management & All \\
Research Exploitation & All \\ 
Training (Workshops, Courses) & All  \\
Dissemination and Communication & All  \\
\bottomrule
\end{tabular}
\end{table}

\paragraph{\textbf{Governance}}
The Certus Center was governed by a board and led by a center director. The board consisted of representatives of the industry partners and the host institution. The Board was meeting bi-annually, discussing the progress of the Certus projects and any matters related to Certus operations.

\paragraph{\textbf{Project Outputs}}
Certus has developed research outputs in the form of scientific publications (journal and conference papers and technical reports), methodology and process guidelines, software prototype tools, and educational material. Examples of different project outputs are shown in Table 1.

Most of the research results achieved in the project had the potential to be developed into industry-relevant results, in the form of software tools or research expertise captured in methodology documents. However, the process of developing such practically relevant results has showed to be challenging without a systematic approach to collaboration.

\subsection{Need for Collaboration Model}
Soon after start-up, we realized that we needed a dedicated collaboration model that will (i) foster research knowledge co-creation through joint problem definition and solving, (ii) promote continuous dialog to help align the expectations of researchers and practitioners, (iii) ensure that the research output created in the project is transformed into results with practical relevance and benefit for the partners, and (iv) ensure that such results are effectively transferred to and exploited by the partners. In addition to creating industrial impact for the project partners, we were also interested in analyzing exploitation opportunities and creating greater impact beyond the project partners for those results that are general and applicable to other target domains. 
Creating a collaboration model supporting these mentioned objectives was the main aim of the governing project Research Exploitation.

\section{Model Development Method}
The Certus collaboration model was developed throughout the project lifetime, incrementally and collaboratively with all project partners. We aimed to allow all project members to shape the collaboration principles. To this end, we used the Participatory Action Research \cite{Reason2008}, which is a knowledge development methodology involving action and reflection. The goal was to provide for everyone to participate in data gathering and analysis. Both data collection and model development was performed in the Research Exploitation governing project, shown in Table 2. The initial version of the model was developed soon after the project start, and then improved in later years. 

\subsection{Data Collection and Analysis}
We used qualitative case study method to collect data for the model development, which included participant observation, reflections, workshops and focus group discussions.
The main method used was participant observation \cite{Taylor1984} \cite{Emerson2001} \cite{DeWalt2010}. This is a method in which data is collected systematically and unobtrusively through social interaction between an observant and informants. The first author of this paper participated both as an observer and active researcher in the collaboration during all 8 years. She was part of one Joint Team in one of the scientific projects, which enabled her to collect various  data during the knowledge co-creation process that might not be noticed otherwise. She was also involved in the Research Exploitation project which allowed her to collect insights on the process of the model development and improvement based on the partners' feedback and data collected in observations.

The data for the model development was collected frequently and in a continuous manner, in a number of occasions. Notes from the meetings and informal discussions among the project participants were taken daily. These notes include information such as motivations for collaboration described by participants, choices of specific methods and technologies for the solution development, the terminology used, choices of evaluation metrics, the types of topics discussed, status updates, and behaviors and interaction patterns between the participants. At the beginning of the project, several workshops and meetings were held with all the partners, dedicated to the question of how the organize effective collaboration between the partners in Certus, during which the observant was taking notes. In addition to the notes from these events, the observer used about 500 emails exchanged between her and other Certus members on the subject of the collaboration, spanning over 8 years.

During data collection, the observer tried to make taking notes as unobtrusive as possible, so that those being observed are not thinking about being observed. This included taking notes electronically, instead of using pen and paper, which appeared natural as everyone in the meeting was usually using a laptop. Immediately after each meeting with note taking, the observer reflected on the information observed and augmented the notes with additional details.

The collected data was analyzed as soon as it was collected. In some cases, the collected data was used to build up the evidence in support of the existing hypotheses about the collaboration, i.e. validating the existing version of the model. In other cases, new hypotheses were defined grounded in the data, for example for extending the model with new aspects. For the latter, we used thematic analysis method \cite{Braun2006} to identify and interpret themes within data. Specifically, the data was first thoroughly read to identify meaningful features in the data, called codes. Some examples of codes include "\textit{goal alignment}", "\textit{problem definition}", "\textit{understanding motivation for research}", "\textit{understanding technical requirements}", "\textit{early interaction between practitioners and researchers}". Coding was done iteratively by combining and splitting codes, and adding new codes. Then we looked for relationships between the codes such as differences and similarities, and grouped the codes to form themes in the data. Next, the themes were reviewed and refined, to make sure they are coherent and accurately represent the data. The theme formed related to the example codes is "\textit{Benefits of intensive interaction between industry and academia at early stage}."

\subsection{Model Development}
The Certus collaboration model was developed in an iterative manner. First, a Working Group was formed consisting of one representative of each project partner. Based on the initial assumptions about the collaboration and the previous experience in industry-academia collaborative projects, the Working Group proposed an initial collaboration model. The initial model was extensively discussed in focus group meetings with each partner individually. These meetings were organised between the representative of Simula, the observer, and relevant staff of a single partner, including engineers and managers. The reason for holding individual meetings with the partners was confidentiality, to allow partners a full freedom to speak, which due to potential conflicts of interest, might not have been achieved if partners had been together in the meeting. The provided feedback was used for the model improvement, to better reflect the common view and expectations from the collaboration. As the scientific projects started activities on problem definition and solution development, additional feedback was provided that served to correct flaws and incorrect assumptions, and to propose a more mature model. The feedback was provided in two ways, by means of: (i) \textit{member checking} \cite{Lincoln1985}, where the collaboration model was presented to the informants at different stages of development, to elicit their validation and comments, and (ii) \textit{observation}, through field notes and reflection of the observer. The feedback was being provided in the same manner throughout the project lifetime, which fed into the model refinement. The illustration of the Certus model development is shown in Figure 1.

\begin{figure}%[htbp]
\centerline{\includegraphics[trim=0cm 11cm 23cm 0cm, width=3.5in]{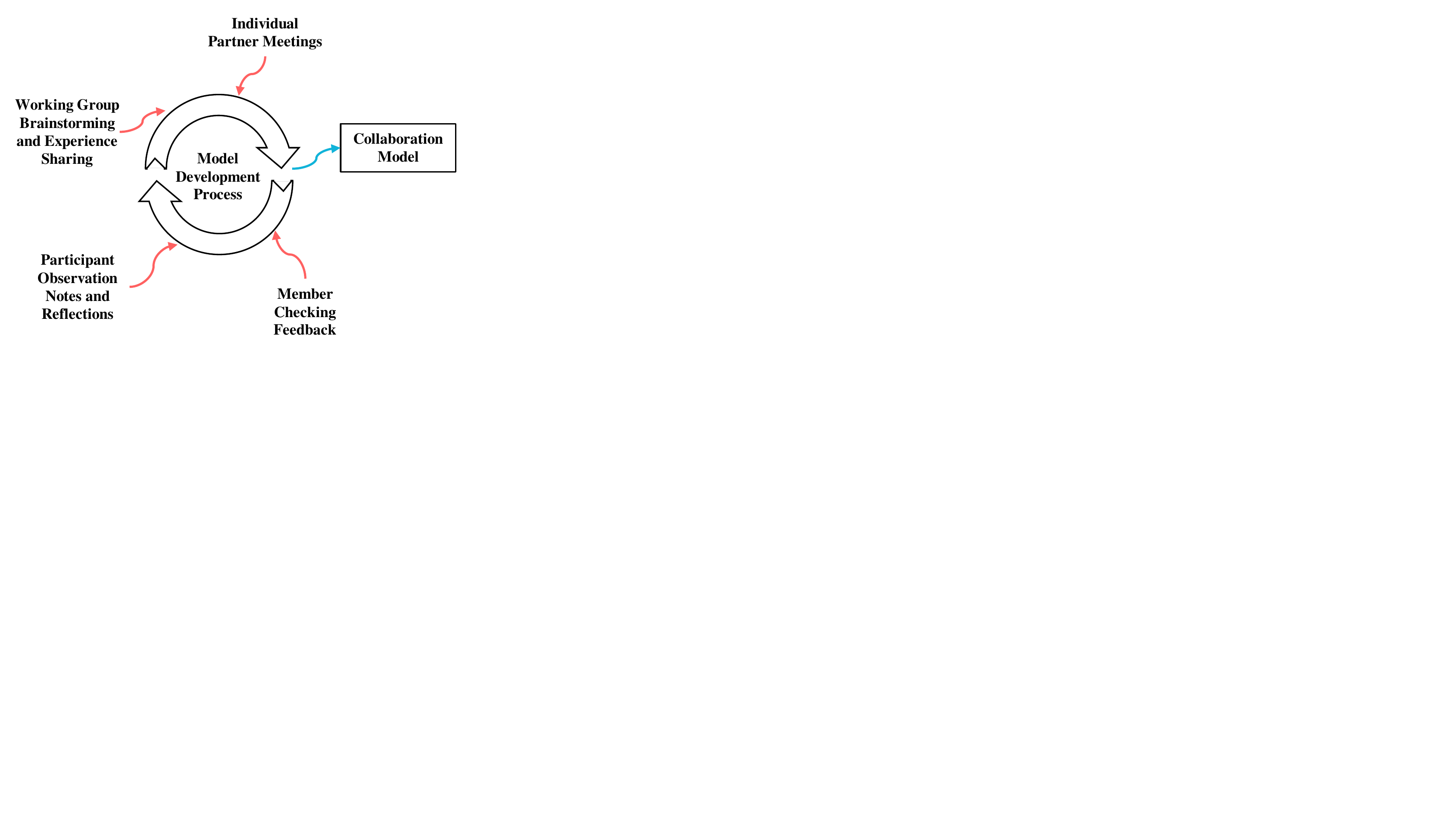}}
\caption{Development process of the Certus Collaboration Model.}
\label{fig}
\end{figure}

\section{Certus Collaboration Model}
In this section, we first present the collaboration setup, followed by the description of the model phases.
 
\subsection{Principles of Collaboration} 
Key ideas underpinning the Certus model include \textbf{\textit{research knowledge co-creation}} and \textbf{\textit{continuous dialog and goal alignment}} between the project participants.   
In the process of research knowledge co-creation, researchers and industry practitioners actively participate in collaboration, by jointly identifying practical (industrial) problems to be solved, translating them into research problems, co-creating solutions and evaluating their potential to solve the practical problems, and reporting the results through research papers. Through experimentation, we realized that co-creation creates the culture of shared ownership, and helps conduct research in a way that makes its results more easily exploitable.

Throughout the collaboration, it is important to enable continuous dialog and goal alignment between researchers and industry practitioners, to keep the problem definition valid, to revisit and revise the partner's expectations, to ensure that the solution being developed meets the technical and non-technical requirements. Dialog helps keep the focus on problem solving and ensures good progress of the work. In addition to enabling active dialog between the members of a single scientific project, we promoted the dialog between the members of different Certus scientific projects, to communicate the progress and results developed among individual projects, which helped identify and build synergies across the Certus Center.

\subsection{Teams and Roles in Collaboration}

\begin{figure}%[htbp]
\centerline{\includegraphics[trim=3cm 1cm 17cm 0cm, width=4.5in]{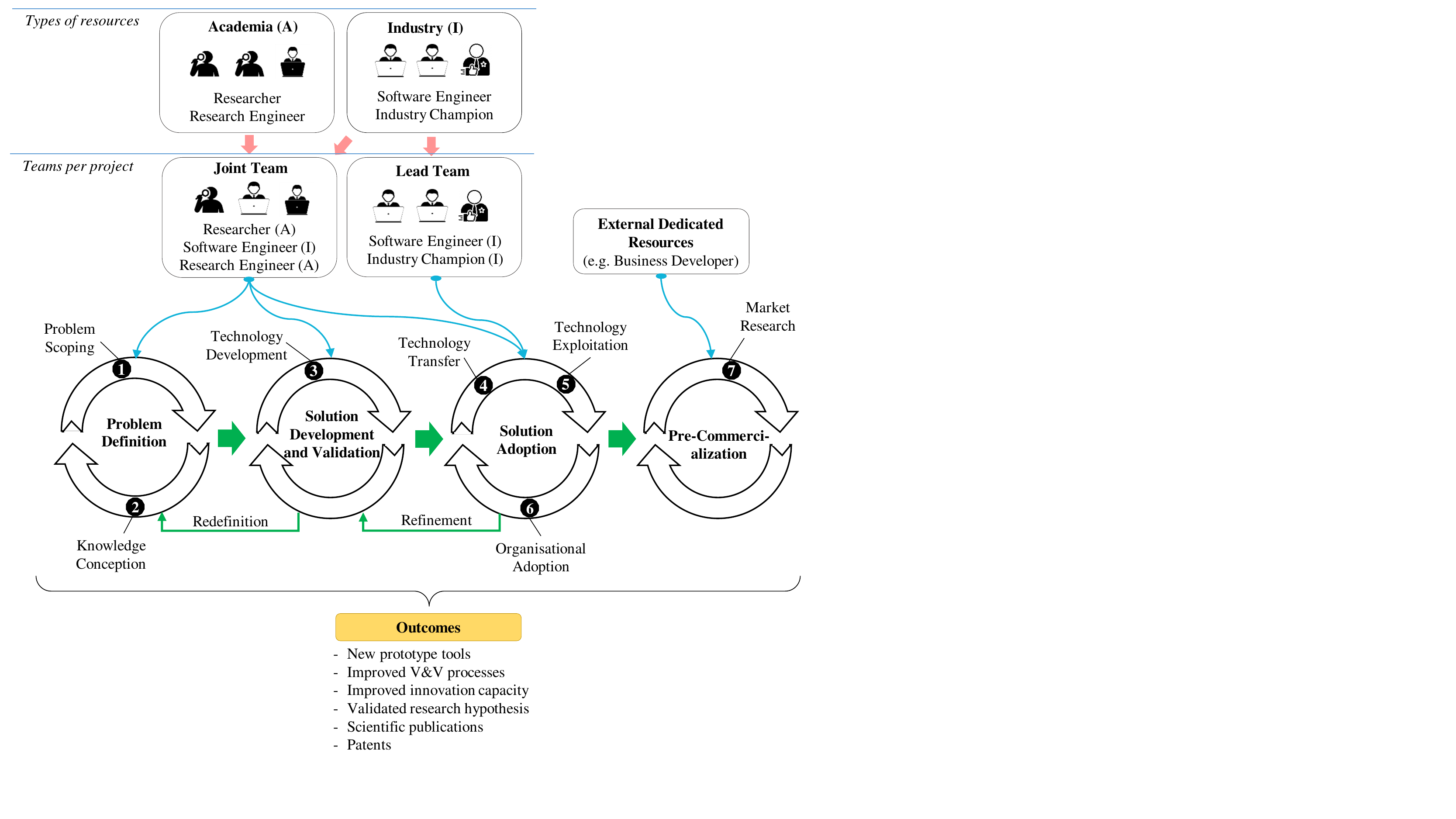}}
\caption{The Joint Team and the Lead Team are exclusive per scientific project. They are composed of different types of resources belonging to academia and industry. The Joint and Lead Team involve in different stages of the collaboration process.}
\label{fig}
\end{figure}

Scientific activities of the Certus Center took place in 6 projects. There is one \textit{\textbf{Joint Team}} formed per project, consisting of 1-2 researchers (including PhD students) and 0-1 research engineers from Simula, and 1-3 industrial software engineers from industry partners' companies, including managers who are typically playing the role of champions. This is illustrated in Figure 2. In principle, Joint Teams are distinct per projects, however, in some cases there is an overlap. This happens when some of the scientific challenges addressed in one project are of interest also for another project. In that case, it is natural to share resources (knowledge, methods, staff) among projects. The staffing of Joint Teams happens after the industrial problems to be solved are defined, such that the right composition of competences can be found. Joint Teams not necessarily need to be fully staffed from beginning, new people can be added/freed when needed. For example, a research engineer's competence is typically required slightly later in the project, after solution concepts have been developed.

At a later stage of technology development, a \textit{\textbf{Lead Team}} is formed, consisting of the partner's engineers and industry champions whose role is to support the technology transfer to the partner's organization, the exploitation of the technology benefits and its wider adoption in the organization. The Lead Team sometimes needs the support of the Joint Team, for example for technology adaptation (see Section 5.7).

\paragraph{\textbf{Day-to-day Operation}}
The research side of the Joint Team works on company sites at times where there is a need for a high level of interaction between the Joint Team members. This happens during the process of industry practice analysis and industry problem definition (see Section 5.4), technology pilot-testing (See Section 5.6), and especially during technology integration with the existing frameworks at the partner's organization, as well as training for technology use (see Section 5.7). At times when researchers perform the state-of-the-art analysis, or devise conceptual solutions, they work in the lab, which provides a more suitable environment for such activities.

\subsection{Model Phases} 
The Certus collaboration model consists of seven phases:  
(1) Problem Scoping, (2) Knowledge Conception, (3) K\&T Development, (4) K\&T Transfer, (5) K\&T Exploitation, (6) Organisational Adoption, and (7) Market Research, graphically presented in Figure 3. 

\begin{figure}%[htbp]
\centerline{\includegraphics[trim=0cm 0cm 15cm 0cm, width=5in]{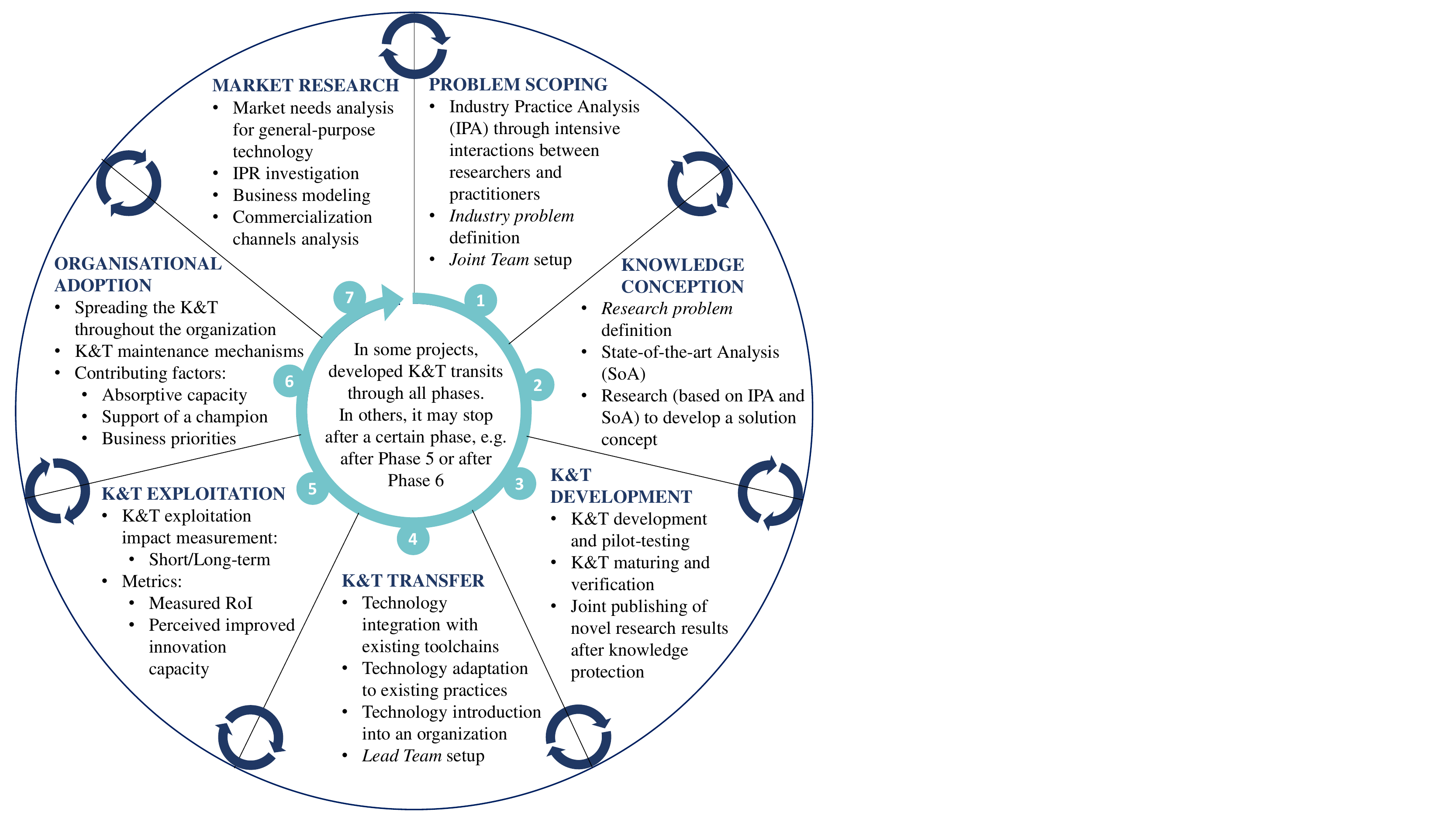}}
\caption{The Certus Collaboration Model.}
\label{fig}
\end{figure}
The Certus model is agile, where outputs of specific phases constantly affect and alter subsequent phases through feedback loops. As a simple example, after a software tool has been introduced in a partner's context and the partner perceives the need for added functionality in the tool transfer phase, this information is fed back into the tool development phase to implement the change, and possibly further to the knowledge conception phase to refine the underlying concepts.

Next we describe each of the model phases.

\subsection{Problem Scoping}
The starting point is a practical or \textit{\textbf{industrial problem}}. The industrial problems we faced in Certus were technical problems arising from underperforming processes in industrial systems. For example, \texttt{high cost of software regression testing}, which is caused by a manual process of test selection. Industrial problem definition is accompanied by an industrial objective, for example, \texttt{to reduce the cost (meaning time in this context) of regression testing by 20\%}.

To identify an industrial problem suitable for addressing collaboratively by researchers and practitioners, there are intensive interactions between the two. Practitioners are describing and demonstrating current practices and their limitations. Researchers are working to understand as much of the domain and business setting as possible, for example, which processes are underperforming, e.g. of low quality or too costly, which tasks and challenges and interrelated, what constraints need to be satisfied. 

Sometimes the defined problem is simpler and can be solved in the short term, and sometimes the problem is complex and needs to be decomposed into smaller sub-problems and tasks which can be prioritized according to urgency, feasibility and risks.

An industrial problem is further restated as a research problem, described in Section 5.5. 

An important element in the initial communication during the problem scoping that has to be understood are expectations for the outcome of the collaboration; what solutions is the partner looking for, what are the technical requirements for the form of the solution sought (e.g., a software tool/service, methodological guidelines), what benefits is the partner expecting and how can those benefits be measured. For example, fewer software field failures that could indicate improved software quality, less time spent in testing or shorter time to market that would indicate improved productivity, or new market opportunities.  
It is especially important to understand what benefit is expected and how it can be measured, because later in the knowledge development stage, the solution is assessed against these metrics. 

After an industrial problem worth pursuing has been defined, a Joint Team is formed to address it.

\subsection{Knowledge Conception}
After an industrial problem has been defined, it is further restated as a \textit{\textbf{research problem}}. A research problem is a statement of a real (industrial) problem that can be solved using research knowledge. Such knowledge can be existing or novel knowledge needs to be generated in the process of problem solving. For our example of the industrial problem given before, a research problem can be stated as the one of \texttt{developing test reduction techniques able to decrease the number of regression tests, such that regression test execution runs 20\% faster, while not decreasing fault detection \\effectiveness of regression tests}.
 
During knowledge conception, a Joint Team is maturing ideas and creating concepts for the problem solution. There are two inputs to this process, as illustrated in Figure 2. One is the analysis of the industry practice, performed in the previous Problem Scoping phase. Another input is the state-of-the-art (SoA) analysis. SoA includes the research solutions, if any, which have been proposed for the same or similar problem, and whether they were applied in practice and how they performed. SoA also includes the analysis of commercial solutions that could solve the problem at hand, and what are the constraints for applying those solutions to the problem. In our experience, it was hardly possible to find an existing commercial solution that can satisfy the constraints of the problem at hand, because the industry domains of our partners were highly specialized and the desired solutions had to be seamlessly integrated with other tools and services. In one case only, we were able to use the existing tool for software variability management, pure::variants \footnote{https://www.pure-systems.com/products/pure-variants-9.html}, as part of the final solution developed, by extending it with a novel research component for automated test selection and reduction.

If no existing solution can be used to adequately address the problem at hand, research activities are started to create novel knowledge/technology building on knowledge gained during the steps of industry practice analysis and SoA analysis. Resulting novel research concepts are analysed by the Joint Team to identify the constraints of the proposed concept solution, to assess its fitness for the problem, and whether it is aligned with the partner's expectations. A decision is made by the Joint Team on whether the conceptual solution has the potential to solve the problem and is worth investing further effort. Knowledge produced in this stage is highly provisional and conceptual, e.g. in the form of presentations, algorithms, etc.

\subsection{Knowledge and Technology Development} 
During this phase, first a prototype tool is developed, or a methodological document is specified following the conceptual solution, then pilot-tested in real setting and matured. Therefore, this phase contains the following two stages.

\paragraph{\textbf{Knowledge and Technology Development}} 
The knowledge underlying the solution concepts is elaborated and reified. If the solution is a particular technology, the knowledge is coded in methods or algorithms, and possibly, developed in the form of a software prototype tool. If the solution comprises practice guidelines, the knowledge is specified through procedures and expressed in the form of methodological documents.
Once the K\&T has been developed (both a software prototype tool and a methodological document), it is assessed for its potential to solve the industrial problem.  
The assessment is performed by the Joint Team, with experiments using real field data %\st{(Milestone 3)}. 
The result of the assessment help decide if the prototype technology should be allocated engineering resources for further technical development and piloting, in which case the Joint Team may be expanded with adequate resources, for example research engineers.

\paragraph{\textbf{Knowledge and Technology Maturing and Verification}} 
For the developed pilot technology, in this stage the first goal is to resolve remaining technical issues and to develop a software tool that is more scalable to the size and complexity of the target domain. A  
\textit{Development Plan} is made to manage the development work in this and the coming Phases, i.e. define activities and responsibilities and allocate resources with a detailed timeline for the expected results, to ensure timely progress of the work. An example of a Development Plan we used in one of the projects contained the following elements: (i) description of the K\&T: its functionality, purpose, areas of application, (ii) partners involved in the development effort (members of the Joint Team and the Lead Team), their roles and responsibilities, (iii) desired main results: software tools, products, processes, services, patents, (iv) detailed activities required to produce the desired results, (v) detailed timeline with milestones for producing the desired results, (vi) resources committed, (vii) intellectual property rights (IPR) aspects, (viii) market analysis for the K\&T to identify technical and market considerations or limitations that can influence technology commercialization, in case the K\&T progresses to commercialization. These elements are added to the Development Plan as they become relevant, i.e. as the K\&T moves through different Phases. A Development Plan is continuously updated as the expectations, goals and motivations of the stakeholders change. 

During the K\&T verification, the newly-developed K\&T is tested in a pilot project, to demonstrate its effectiveness (i.e., problem-solving capabilities for a smaller-size problem) and to assess its potential to succeed in a full-scale project. The assessment is performed by the Joint Team using real datasets and measuring against the metrics achieved by current industry practice.  
The result of the assessment helps determine if the developed K\&T has the potential for successful exploitation in the partner's organization and is thus a candidate for the K\&T transfer. 

Furthermore, the results of the K\&T verification help prove or disprove the hypothesis underlying the research problems defined in Phase 2. This creates the opportunity for publishing novel research results as technical contributions (algorithms, tools), or empirical research (case studies, experience reports). Often, as it was the case in our projects, researchers and practitioners jointly co-author papers. A good practice is to consider knowledge protection needs before knowledge publishing. Such considerations are made by the Joint Team, sometimes involving higher management, to decide whether industry partners have any interests in protecting the knowledge before publishing.

\subsection{Knowledge and Technology Transfer}
The goal of this phase is to put the partner in full possession of the verified K\&T and to create the conditions such that the K\&T can be applied and maintained. In practical terms, this means that the partner can independently use the K\&T as part of the existing tools and processes and is able to spread it further within the organization. In this phase a Lead Team is formed. We distinguish two following stages in this process.

\paragraph{\textbf{Technology Integration and Adaptation}}
This is mainly an engineering stage, required for transferring the technology in the form of software tools (i.e., it is not applicable for knowledge in the form of guideline documents). Technology being transferred is integrated with the partner's existing tools and processes. Technical integration usually requires some adaptation of the tool, e.g. interfacing with the existing tools, and this work is done jointly by the Joint Team and the Lead Team. After the solution has been made usable, additional effort is needed to make it fully adopted by the partner and used without any assistance from researchers.

\paragraph{\textbf{Knowledge and Technology Introduction}}
Introducing new knowledge and technology into an organization presents a set of challenges, and a key to success is starting K\&T introduction planning sufficiently early, along with the K\&T technical integration. The goal of this stage is to prepare the partner organization for receiving the K\&T and make a plan for the hand-off. 
First, a Lead Team receives training to gain full ownership over the K\&T. The engineers who are part of the Joint Team already own the underlying concepts. The goal now is to make a larger group of partner's engineers familiar with these concepts and able to use the K\&T effortlessly and independently in different scenarios, as well as to propagate it throughout the organization. The K\&T introduction phase includes the training for K\&T use, given by researchers and supported by reference manuals for the new technology. 
Besides the training for use, there is training for technology maintenance. Maintenance can be required when new requirements are raised, or when parts of the system interacting with the transferred technology change. For some partners, maintenance can be done in-house and in that case engineers should be trained to support maintenance activities. For the partners who do not have the resources to support maintenance, there are several mechanisms that can be considered for maintaining transferred technology, discussed in Section 5.9. The K\&T introduction stage only makes plans for technology maintenance, while the training for maintenance takes place in the K\&T adoption process.

\subsection{Knowledge and Technology Exploitation}
After the K\&T has been transferred to the partner's organization and the Lead Team has been able to apply it, it is sometimes possible to quantify the economic benefit gained from the research investment and, thus, calculate the K\&T exploitation impact. The partner can start exploiting the K\&T before it is widely adopted in the organization, since the adoption process is slow-paced and happens gradually. Besides, some K\&T may only make a sizeable impact in a particular sub-unit rather than the enterprise as a whole.  
The exploitation impact of a transferred and introduced K\&T is correlated with the degree of economic gain that the K\&T has made for the partner. In the context of Certus, examples of such gains include reduced costs of software testing, increased revenue due to better product quality and more sales, or new business opportunities. The economic impact of a new K\&T can be achieved either in the short or the long term. Some types of K\&T have short lead time in demonstrating their impact and bringing economic gain, such as, for example, new software tools. Other K\&T enhance the partners' ability to innovate through increased knowledge and understanding of specific research topics, thus creating an incremental impact by improving organizational innovation capacity and performance in the long run. For the Certus projects, it was sometimes possible to approximate the K\&T exploitation impact using two metrics that measure both a short-term and long-term economic impact. However, we acknowledge the challenge of measuring the exact impact of a newly introduced technology, because such technology does not operate in a vacuum, but instead there are factors coming from surrounding technologies or processes that evolve and improve themselves, and that can fully or partly affect the new technology operation. The two metrics we used in Certus projects include:
\begin{itemize}
\item[--] RoI, calculated as \texttt{(gain from investment - cost of investment)/\\cost of investment}, as a quantitative metric. Gain from investment is the financial gain obtained by using a novel K\&T technology, for example, reducing the cost of regression testing by 20\% (our previous example). Cost of investment can include the cost of transferring the K\&T into the partner's organization (engineering effort of integrating and adapting the solution), training costs (effort of learning new skills), operating costs (effort of using K\&T), maintenance costs, costs of licenses as part of the K\&T. However, accurately calculating the benefit of investment is considered much harder then calculating the cost of investment \cite{Punter2006}.
\item[--] Partner's perception of the benefit obtained in terms of strengthened innovation capacity, as a qualitative metric collected through interviews and questionnaires.  
\end{itemize}

\subsection{Organisational Adoption}
K\&T organizational adoption is a process where a Lead Team as the owner of the K\&T spreads the technology throughout the organisation into routine use.  
A Joint Team assists the Lead Team, if necessary, and provides training for the technology maintenance. There could be several maintenance mechanism, depending on the partner's motivation, skills and available resources, as follows: (i) the partner allocates resources to maintain the technology in-house, (ii) if there is a tool vendor partner among the project partners, this partner analyses their motivation and competence to provide maintenance services for the technology and to explore possible commercialization opportunities. In that case researchers help the tool vendor partner understand the technology. IPR also needs to be investigated, (iii) the technology maintenance is outsourced to a third party. IPR needs to be investigated as well, (iv) the technology maintenance is handled through a contract research service by the host research institute.

K\&T adoption process is known to be difficult and slow \cite{Redwine1984}, taking even 10 up to 15 years \cite{Pfleeger1999} and therefore needs to be supported by comprehensive plans, as it requires altering or even discarding old practices and employing and sustaining new ones. One important factor that influences the adoption process is the organizational absorptive capacity \cite{Rajalo2017}, as the ability of the partner to %\st{absorb} 
acquire and assimilate new research-based K\&T. Absorptive capacity has been considered as an important contributor to the success of technology transfer in academic engagements \cite{Vries2019}. Organizations with stronger absorptive capacity will more readily recognize the value and accept a new K\&T. Since the level of absorptive capacity partly depends on prior knowledge and experience, as well as innovation capacity, the development of innovation capacity of the Certus partners is considered an important outcome of the Certus Center. %, creating a long-term economic impact for the partners. 
Furthermore, an important factor that negatively influences the adoption process is discontinuity of the partner's business focus and priorities, where the organizational strategy becomes no longer aligned with the investment in the area concerned with the K\&T that is a candidate for adoption. The same applies to cases where K\&T champions leave the partner's organization. In that case, in our experience, the co-creation process typically slows down and the dialog becomes more sparse. Another factors that affect the adoption process include the organization's financial resilience and maintenance of the innovation capacity.

\subsection{Market Research}
Some of the research results produced in a collaborative project between industry and academia may have the potential for achieving greater benefits outside the context of the collaborative project. These could be, for example, general-purpose technology that has a broad spectrum of applications, and for which the industry partners do not have interest to pursue commercialization themselves. In such cases, if a host research institute has the resources, they can investigate the market potential for these value propositions, and examine the most appropriate channels for their commercialization. The motivation of the research institute for doing such type of activities includes potentially increased financial portfolio, as well as improved visibility and innovation reputation, in case that the technology proves the market potential and that its commercialization is successful.     

Factors to be considered in an initial market research include: (i) the existence of similar and competing technologies, (ii) novelty and added value compared to these technologies, (iii) customer interest in the value proposition and the level of their interest, e.g. willingness to invest in further development, adaptation and transfer, (iv) limitations that can influence commercialization, e.g. weaknesses of the value proposition, (v) commercialization channels, e.g. spin-offs for product development or consultancy services, or technology licensing, or technology selling, (vi) resources and partners needed for commercialization, and (vii) comprehensive IPR investigation. If the initial analysis shows the market potential for the value proposition, a more systematic analysis should be performed. This may include the development of a business model, to test the feasibility of the business idea for the value proposition. The business model helps identify target customers for the technology, examine the infrastructure needed for developing a commercial product from the technology, and analyze cost and revenue structures. If a detailed analysis shows potential for the technology commercialization, actual commercialization, as well as IP protection (if needed), needs to be pursued by other means than the collaboration project. Such means may include the creation of a spin-off, or technology licensing or selling, since technology commercialization activities exceed the scope and resources of the collaboration project.

In case of a spin-off creation, in our experience, it is recommended to investigate several aspects: (i) are there highly motivated and skilled people to set on a journey of company creation. For example, it is beneficial to have on-board a researcher who had a key role in the technology creation. However, it needs to be confirmed whether s(he) has aspirations for entrepreneurship or (s)he wants to continue a research career, (ii) what support and incentives are provided by the research institute for a researcher leaving their research position. Some employees of early-stage companies make no salary before the company has increased valuation. In that case, keeping a part-time research position provides some level of financial security for a researcher until the spin-off company reduces risks and increases value, (iii) plans for company financing.  

Analysing the market potential of a research-based technology usually falls outside the job scope of a typical researcher. In the case of Certus researchers, the host research institute established a dedicated project with suitable resources for dealing with the market research and pre-commercialization activities for one of the research-based technologies developed in one of the scientific projects.

\section{Certus Model Evolution: Example with Cisco Systems}
In this section we describe the life of one collaborative project guided by the Certus collaboration model. This was a scientific project \textit{Testing of Real Time Embedded Systems} in collaboration with Cisco Systems. We give examples of the model evolution based on the experiences made and knowledge generated during the project execution. 

At the project beginning, Cisco test engineers revealed the problem of high test effort in regression testing they are facing, coming from (semi-)manual test selection. The software under test was highly configurable, which means that it could be run with a large number of varying input parameters and their combinations. This created a need for effective management of test complexity caused by variations in system structure and functionality. After a set of initial discussions about the practical Cisco problem, a Joint Team was formed consisting of 3 test engineers from Cisco, 2 research scientist and 1 PhD Student from Simula. The Joint Team started the problem scoping by observing the industrial practice of testing highly configurable systems, with the goal of identifying addressable software testing challenges. This initial phase of challenge identification took place in the form of frequent one-on-one meetings and technical workshops, where researchers were learning domain specifics necessary for problem formulation. Furthermore, a common terminology was built, which was useful throughout the whole project. The industry problem was defined as the one of \textit{reducing the cost of regression testing by 20\%}. After analyzing existing approaches to such similar problems, we found that effective testing in such an environment aimed for automated and optimized mechanisms to reduce manual effort, and improve the testing process workflow. The research problem was defined as the one of \textit{developing test reduction techniques able to reduce the number of regression tests such that regression testing runs 20\% faster}. Based on these objectives, the Joint Team started with the concept solution development. The K\&T Development stage took the Joint Team around two man-years to develop a candidate proof-of-concept that was experimentally tested and improved based on the feedback from engineers. There were a lot of iterations between the phases of Knowledge Conception and K\&T Development, for refining the solution concept, based on the feedback from the solution development and the insights from the evolving state-of-the-art. The Joint Team tried to make demonstrable progres all the time, which helped keep all team members enthusiastic. The solution built to address the industrial problem was TITAN \cite{Marijan2017}, a software testing tool that automates and optimizes several stages in software testing processes, balancing the expenditure and quality of testing \cite{6676952, 7272927, Marijan2019, 7816510, 8377636, 3183532, 7911878}. TITAN development was an iterative process where the  solution was validated in an operational environment, using real industrial data and artifacts, and then improved following the feedback given in validation. The Joint Team evaluated TITAN with respect to the time reduction metric of 20\%. After successful evaluation, the project transitioned to the K\&T Transfer phase. In this phase the Lead Team was formed, who worked with the Joint Team to integrate TITAN and adapt it to the existing testing framework. In this phase, it became evident that modifications need to be made in the technical implementation of TITAN to enable seamless integration. TITAN was an Eclipse plugin, while Cisco needed it to be a software service for a smooth integration. As the Cisco test engineer said "\textit{Unless we are able to smoothly integrate the tool, there is not much chance our engineers will use it".} This feedback was propagated to the K\&T Development phase where changes were made in the technical choices for the TITAN solution. This is an example of the experience from one project being used to update the model and provide a lesson learned for other projects. The \textbf{learning} is that listening to and understanding the technical requirements for the solution being developed is crucial before the technical development starts, in order for the solution to be useful in practice. At that point, the Joint Team was extended by one research engineer from Simula, whose expertise was suitable for the type of work needed. Based on the feedback, the Joint Team continued with the technology development and experimentation in the K\&T phase, until the technical solution was produced that can easily integrate with the existing testing framework at Cisco, in the K\&T Transfer phase. After that, the Lead Team tried to measure the benefit of TITAN. Precise quantitative measurement was challenging, because there were different factors affecting the overall test time in regression testing. However, in qualitative assessment, the industrial part of the Joint Team and the Lead Team stated that working on the project with Simula researchers improved their understanding of particular research techniques on software testing and increased their ability to integrate research results into their practice. The Certus model refers to this type of impact as \textit{improved innovation capacity}. In the Organizational Adoption phase of TITAN, there was an industry champion at Cisco who played the key role in the tool adoption. He was open-minded and passionate about research and technology, and he was a member of the Lead Team. He was at the middle-management level and was actively promoting the tool among other colleagues. This resulted in TITAN being adopted within one team, who dealt with software quality assurance. TITAN was not spread throughout the entire organization due to the nature of that Cisco business unit. The purpose of the tool was to improve regression testing, which was not central to the core business units dealing with video communication technology. Further, based on the interaction with tool vendors outside of the Certus projects, we hypothesized that TITAN could be useful in other domains in need of test optimization and therefore started to work on Market Research for the tool. After the initial analysis proved promising, Simula obtained additional funding aimed at testing the market potential for TITAN. A business model was developed proving the commercial potential of the tool, and together with external resources (business developers and advisers) we agreed that the most appropriate commercialization channel is a spin-off. One of the business advisers advised that \textit{"It is very important that at least one of the technology inventors joins the company. This is because in a technology company it is critical to master the technology in early stages before the first sales are obtained and technology proved useful for customers"}. At that point, discussions started on clarifying the IPR and the incentives of key researchers for joining the spin-off. However, the discussions revealed that the key researchers involved in TITAN development did not have aspirations for entrepreneurship. This is another example where the experience from one project improved the Certus model. A \textbf{lesson learned} here is that the motivations of key researchers for joining the spin-off should be clarified early on. However, during the Market Research phase, ideas were generated for new projects that were later addressed together with the Certus industry partners and other external partners, feeding into the Problem Scoping phase.

\section{Discussion and Conclusion}
This paper presents the Certus collaboration model, which answers the research question of what elements are needed for an industry-academia collaboration to support research knowledge co-creation in SE. These elements are problem scoping, knowledge conception, K\&T development, K\&T transfer, K\&T exploitation, organisational adoption, and market research. In our experience, a collaboration model supporting research knowledge co-creation has the potential to overcome notable collaboration challenges such as the lack of relevance, interest, and commitment from the collaboration parties. Compared to other industry-academia collaboration models proposed in SE, the Certus model presents a practical collaboration framework that details different roles and responsibilities of project participants, ways of promoting commitment and participative knowledge generation among project participants, as well as some directions towards the commercialization of research results.

\subsection{Reflections on Model Development}
The model was developed iteratively, embodying only preliminary elements of collaboration at beginning, and gradually evolving as the collaboration progressed and as we gained more experience of co-creation. The very initial version of the model did not promote the concept of co-creation early enough. For example, a Joint Team was not formed already in the Problem Scoping phase, but more towards the K\&T Development phase. This resulted in a more loose ownership of the problem definition by industry partners, and consequently less active participation in concept development. After the benefit of forming a Joint Team during a Problem Scoping phase became evident in one of the scientific projects, the model was improved.

Even if all model phases posed certain challenges for the model development, there was one part which was the most challenging. That was devising the metrics that can be used for measuring the K\&T exploitation impact. For most of our projects, it came down to measuring the impact of a new tool that works as part of a bigger toolchain that evolves and improves performance itself overtime. As a quantitative metrics, we proposed to approximate RoI, which was sometimes feasible. More feasible was to analyze an improved innovation capacity, as a qualitative metric, through interviews and focus group discussions.
  
\subsection{Limitations}
There are several limitations and validity threats of the Certus model.

\textit{Internal Validity.} The collection of data underpinning the Certus model relies on the participant observation method, which has several limitations. First, the method is subjective as it represents a perspective of an individual observer. To mitigate this threat, triangulation of three data sources was used, which included emails, field notes, and feedback from the informants, both the Working Group members and the Joint Teams on their perspective on the collaboration model. Feedback from the Working Group and Joint Teams was provided by means of the member checking technique, where the collaboration model was presented to the informants at different stages of development to obtain their input and support for further model development. Member checking is considered especially useful in our case because the findings made in observation may change the collaboration model which further changes the way the Certus partners collaborate \cite{Seaman1999}. Second threat to validity is representativeness, which means that informants may not be representative of all relevant participants in the collaboration. To mitigate this threat, we conducted participant observation in different settings involving participants with different experience and expertise. Software engineers and researchers were observed as part of Joint Teams, managers were observed in focus group discussions and workshops. Third threat to validity is the observer-expectancy effect, which happens when the observer's presence affects the informants' behavior. To mitigate this threat, the observer was taking notes on a laptop, trying to be as unobtrusive as possible. 

\textit{External Validity.} Our findings on industry-academia collaboration underpinning the Certus collaboration model are based on one specific collaborative setting with a specific
funding and governing scheme, untested in other environments, which limits the generalizability of our findings. Further work is necessary to understand how well our model generalizes to other industry-academia collaborative settings and how it could be adapted to other contexts.

\subsection{Further Adaptation and Application}
As already pointed out, we do not know how well the Certus model could generalize to another collaboration context and what model modification would need to be made. Just like we experimented with its initial version and improved it overtime, applying the model to another collaboration context would be similar. It would take starting from the current version and adding additional elements on demand. Which elements would be needed depends on the context. Even more, some of the elements available in the current model might not be appropriate for a particular context and would be discontinued.    
  
\section*{Acknowledgment}
This work is supported by the Research Council of Norway, through the Certus SFI project.

\section*{References}
\bibliographystyle{elsarticle-num}
\bibliography{mybibfile}

\begin{thebibliography}{10}
\expandafter\ifx\csname url\endcsname\relax
  \def\url#1{\texttt{#1}}\fi
\expandafter\ifx\csname urlprefix\endcsname\relax\def\urlprefix{URL }\fi
\expandafter\ifx\csname href\endcsname\relax
  \def\href#1#2{#2} \def\path#1{#1}\fi

\bibitem{Wirsich2016}
A.~Wirsich, A.~Kock, C.~Strumann, C.~Schultz, Effects of university-industry
  collaboration on technological newness of firms, Journal of Product
  Innovation Management 33~(6) (2016) 708--725.

\bibitem{Perkmann2010}
M.~Perkmann, V.~Tartari, M.~McKelvey, E.~Autio, et. al, Academia engagement and
  commercialization: A review of the literature on university-industry
  relations, Research Policy 42 (2010) 423--442.

\bibitem{Lamprecht2012}
S.~Lamprecht, G.~Rooyen, Models for technology research collaboration between
  industry and academia in south africa, IEEE Software Engineering Colloquium
  (2012) 11--17.

\bibitem{Bosch2014}
J.~Bosch, Continuous software engineering: An introduction, Continuous Software
  Engineering, Springer, Cham. (2014) 3--13\href
  {http://dx.doi.org/10.1007/978-3-319-11283-1}
  {\path{doi:10.1007/978-3-319-11283-1}}.

\bibitem{Chimalakonda2015}
S.~Chimalakonda, Y.~Reddy, R.~Shukla, Moving beyond: Insights from 1st
  international workshop on software engineering research and industrial
  practices, ACM SIGSOFT Software Engineering Notes (2015) 28--31\href
  {http://dx.doi.org/10.1145/2735399.2735418}
  {\path{doi:10.1145/2735399.2735418}}.

\bibitem{Runeson2014a}
P.~Runeson, S.~Minor, J.~Svener, Get the cogs in synch – time horizon aspects
  of industry–academia collaboration, ACM International Workshop on Long-term
  Industrial Collaboration on Software Engineering 2014 (2014) 25--28.

\bibitem{Barroca2015}
L.~Barroca, H.~Sharp, D.~Salah, K.~Taylor, P.~Gregory, Bridging the gap between
  research and agile practice: an evolutionary model, International Journal of
  System Assurance Engineering and Management 9 (2015) 323--334.

\bibitem{Ivanov2017}
V.~Ivanov, A.~Rogers, G.~Succi, Y.~Jooyong, V.~Zorin, What do software
  engineers care about: gaps between research and practice, Joint Meeting on
  Foundations of Software Engineering (2017) 890--895.

\bibitem{Osterweil2008}
L.~J. Osterweil, C.~Ghezzi, J.~Kramer, A.~L. Wolf, Determining the impact of
  software engineering research on practice, Computer 41~(3) (2008) 39--49.

\bibitem{Emmerich2005}
W.~Emmerich, M.~Aoyama, J.~Sventek, Impact of research on the development of
  middleware technology, ACM TOSEM 17~(4) (2005) 1--48.

\bibitem{Icse2011}
ICSE, Icse 2011 panel on what industry wants from research, ICSE, Hawaii, USA
  (2011).

\bibitem{Briand2012}
L.~Briand, Embracing the engineering side of software engineering, IEEE
  Software 29~(4) (2012) 96--96.

\bibitem{Briand2017}
L.~Briand, D.~Bianculli, S.~Nejati, F.~Pastore, M.~Sabetzadeh, The case for
  context-driven software engineering research: Generalizability is overrated,
  IEEE Software 34~(5) (2017) 72--75.

\bibitem{Bern2018}
B.~Bern, From theory to practice: Experiences of industry-academia
  collaboration from a practitioner, IEEE/ACM International Workshop on Sftware
  Engineering Research and Industrial Practice 2018 (2018) 22--23.

\bibitem{Tang2017}
A.~Tang, R.~Kazman, On the worthiness of software engineering research,
  Technical report (2017).

\bibitem{Garousi2018}
V.~Garousi, M.~Borg, M.~Kuhrmann, M.~Oivo, Cut to the chase: Revisiting the
  relevance of software engineering research, Int. Conference on Evaluation and
  Assessment in Software Engineering 2018 25 (2018) 1687--1754.

\bibitem{Basili2018}
V.~Basili, L.~Briand, D.~Bianculli, S.~Nejati, F.~Pastore, M.~Sabetzadeh,
  Software engineering research and industry: a symbiotic relationship to
  foster impact, IEEE Software 35~(5) (2018) 44--49.

\bibitem{Carver2018}
J.~Carver, R.~Prikladnicki, Industry-academia collaboration in software
  engineering, IEEE Software 35~(5) (2018) 120--124.

\bibitem{Garousi2016}
V.~Garousi, K.~Petersen, B.~Ozkan, Challenges and best practices in
  industry-academia collaborations in software engineering: a systematic
  literature review, Information and Software Technology 79 (2016) 106--127.

\bibitem{Garousi2017a}
V.~Garousi, M.~Felderer, J.~Fernandes, D.~Pfahl, M.~Mantyla, Industry-academia
  collaborations in software engineering: an empirical analysis of challenges,
  patterns, and anti-patterns in research projects, Int. Conference on
  Evaluation and Assessment in Software Engineering (2017) 224--229.

\bibitem{Pertuze2010}
J.~Pertuze, E.~Calder, E.~Greitzer, W.~Lucas, Best practices for
  industry-university collaboration, MITSloan Management Review (2010).

\bibitem{Wohlin2011}
C.~Wohlin, A.~Aurum, L.~Angelis, L.~Phillips, Y.~Dittrich, T.~Gorschek, Success
  factors powering industry-academia collaboration, IEEE Software 29~(2) (2011)
  67--73.

\bibitem{Wohlin2013}
C.~Wohlin, Software engineering research under the lamppost, International
  Joint Conference on Software Technologies (2013) IS-11.

\bibitem{Petersen2014}
K.~Petersen, C.~Gencel, N.~Asghari, D.~Baca, S.~Betz, Action research as a
  model for industry-academia collaboration in the software engineering
  context, ACM Int. Workshop on Long-term Industrial Collaboration on Software
  Engineering (2014) 55--62.

\bibitem{Sandberg2011}
A.~Sandberg, L.~Pareto, T.~Arts, Agile collaborative research: action
  principles for industry-academia collaboration, IEEE Software 28~(4) (2011)
  74--83.

\bibitem{Sjoo2019}
K.~Sjoo, T.~Hellstrom, University-industry collaboration: A literature review
  and synthesis, Industry and Higher Education 33~(4) (2001) 275--285.

\bibitem{Dittrich2008}
Y.~Dittrich, K.~Rönkkö, J.~Eriksson, C.~Hansson, O.~Lindeberg, Cooperative
  method development: Combining qualitative empirical research with method,
  technique and process improvement, Empirical Software Engineering 13 (2008)
  231--260.

\bibitem{Mathiassen2012}
L.~Mathiassen, Collaborative practice research, Information, Technology and
  People 14 (2012) 321--345.

\bibitem{Lin2017}
J.~Lin, Balancing industry collaboration and academic innovation: the contigent
  role of collaboration specific attributes, Technological Forecast and Social
  Change 123 (2017) 216--228.

\bibitem{Huang2017}
M.~Huang, D.~Chen, How can academic innovation perofrmance in
  university-industry collaboration be improved, Technological Forecast and
  Social Change 123 (2017) 210--215.

\bibitem{Banal2013}
A.~B. Estanol, I.~M. Stadler, D.~P. Castrillo, Research output from
  university-industry collaborative projects, Economic Development 27~(1)
  (2013) 71--81.

\bibitem{Runeson2014}
P.~Runeson, S.~Minor, The 4+1 view model of industry-academia collaboration,
  ACM International Workshop on Long-term Industrial Collaboration on Software
  Engineering (2014) 21--24.

\bibitem{Garousi2019}
V.~Garousi, D.~Pfahl, J.~Fernandes, M.~Felderer, M.~Mantyla, D.~Shepherd,
  Characterizing industry-academia collaborations in software engineering:
  evidence from 101 projects, Empirical Software Engineering 24 (2019)
  2540--2602.

\bibitem{Gorschek2006}
T.~Gorschek, P.~Garre, S.~Larson, C.~Wohlin, A model for technology transfer in
  practice, IEEE Software 23~(6) (2006) 88--95.

\bibitem{Pfleeger1999}
S.~Pfleeger, Understanding and improving technogloy transfer in software
  engineering, J. Systems and Software 47~(3) (1999) 111--124.

\bibitem{Reason2008}
P.~Reason, H.~Bradbury, The sage handbook of action research: Participative
  inquiry and practice, Sage, CA, Second Edition (2008).

\bibitem{Taylor1984}
S.~Taylor, R.~Bogdan, Introduction to qualitative research methods, New York:
  John Wiley and Sons (1984).

\bibitem{Emerson2001}
R.~Emerson, R.~Fretz, L.~Shaw, Participant observation and fieldnotes, Handbook
  of Ethnography (2001) 356--357.

\bibitem{DeWalt2010}
K.~DeWalt, B.~DeWalt, C.~Wayland, Participant observation: A guide for
  fieldworkers, AltaMira Press (2010), ISBN 0759100446.

\bibitem{Braun2006}
V.~Braun, V.~Clarke, Using thematic analysis in psychology, Qualitative
  Research in Psychology 3 (2006) 77--101.

\bibitem{Lincoln1985}
Y.~Lincoln, E.~Guba, Naturalistic inquiry, Thousand Oaks Calif.: Sage (1985).

\bibitem{Punter2006}
T.~Punter, R.~Krikhaar, R.~Bril, Sustainable technology transfer, International
  Workshop on Software Technology Transfer in Software Engineering (2006)
  15--18.

\bibitem{Redwine1984}
S.~Redwine, W.~Riddle, Software technology maturation, International Conference
  on Software Engineering (1984) 189--200.

\bibitem{Rajalo2017}
S.~Rajalo, M.~Vadi, University-industry innovation and collaboration:
  Reconceptualization, Technovation 62 (2017) 42--54.

\bibitem{Vries2019}
E.~Vries, W.~Dolfsma, M.~Gerkema, Knowledge transfer in university-industry
  research partnerships: a review, The Journal of Technology Transfer 44 (2019)
  1236--1255.

\bibitem{Marijan2017}
D.~Marijan, M.~Liaaen, A.~Gotlieb, S.~Sen, C.~Ieva, Titan: Test suite
  optimization for highly configurable software, International Conference on
  Software Testing (2017) 524--531.

\bibitem{6676952}
D.~Marijan, A.~Gotlieb, S.~Sen, Test case prioritization for continuous
  regression testing: An industrial case study, in: 2013 IEEE International
  Conference on Software Maintenance, 2013, pp. 540--543.
\newblock \href {http://dx.doi.org/10.1109/ICSM.2013.91}
  {\path{doi:10.1109/ICSM.2013.91}}.

\bibitem{7272927}
D.~Marijan, Multi-perspective regression test prioritization for
  time-constrained environments, in: 2015 IEEE International Conference on
  Software Quality, Reliability and Security, 2015, pp. 157--162.
\newblock \href {http://dx.doi.org/10.1109/QRS.2015.31}
  {\path{doi:10.1109/QRS.2015.31}}.

\bibitem{Marijan2019}
D.~Marijan, A.~Gotlieb, M.~Liaaen, A learning algorithm for optimizing
  continuous integration development and testing practice, in: Software:
  Practice and Experience, 2019, pp. 192--213.
\newblock \href {http://dx.doi.org/doi.org/10.1002/spe.2661}
  {\path{doi:doi.org/10.1002/spe.2661}}.

\bibitem{7816510}
D.~Marijan, M.~Liaaen, Effect of time window on the performance of continuous
  regression testing, in: 2016 IEEE International Conference on Software
  Maintenance and Evolution (ICSME), 2016, pp. 568--571.
\newblock \href {http://dx.doi.org/10.1109/ICSME.2016.77}
  {\path{doi:10.1109/ICSME.2016.77}}.

\bibitem{8377636}
D.~Marijan, M.~Liaaen, S.~Sen, Devops improvements for reduced cycle times with
  integrated test optimizations for continuous integration, in: 2018 IEEE 42nd
  Annual Computer Software and Applications Conference (COMPSAC), Vol.~01,
  2018, pp. 22--27.
\newblock \href {http://dx.doi.org/10.1109/COMPSAC.2018.00012}
  {\path{doi:10.1109/COMPSAC.2018.00012}}.

\bibitem{3183532}
D.~Marijan, M.~Liaaen, \href{https://doi.org/10.1145/3183519.3183532}{Practical
  selective regression testing with effective redundancy in interleaved tests},
  in: Proceedings of the 40th International Conference on Software Engineering:
  Software Engineering in Practice, ICSE-SEIP '18, Association for Computing
  Machinery, New York, NY, USA, 2018, p. 153–162.
\newblock \href {http://dx.doi.org/10.1145/3183519.3183532}
  {\path{doi:10.1145/3183519.3183532}}.
\newline\urlprefix\url{https://doi.org/10.1145/3183519.3183532}

\bibitem{7911878}
D.~Marijan, M.~Liaaen, Test prioritization with optimally balanced
  configuration coverage, in: 2017 IEEE 18th International Symposium on High
  Assurance Systems Engineering (HASE), 2017, pp. 100--103.
\newblock \href {http://dx.doi.org/10.1109/HASE.2017.26}
  {\path{doi:10.1109/HASE.2017.26}}.

\bibitem{Seaman1999}
C.~Seaman, Qualitative methods in empirical studies of software engineering,
  IEEE Transactions on Software Engineering 25 (1999) 557--572.

\end{thebibliography}

\end{document}